\documentclass[preprint,preprintnumbers,amsmath,amssymb]{revtex4}

\usepackage{graphicx}
\usepackage{dcolumn}
\usepackage{bm}
\usepackage{txfonts}

\begin{document}

\preprint{}
\title{Generation of arbitrary full Poincar\'{e} beams on the hybrid-order Poincar\'{e} sphere}
\author{Xiaohui Ling,$^{1,2}$ }\email{xhling@hnu.edu.cn}
\author{Xunong Yi$^{1}$}
\author{Zhiping Dai$^{2}$}
\author{Youwen Wang$^{2}$}
\author{Liezun Chen$^{2}$}

\address{$^1$SZU-NUS Collaborative Innovation Center for Optoelectronic Science and Technology, and Key Laboratory of Optoelectronic Devices and
Systems of Ministry of Education and Guangdong Province, College of Optoelectronic Engineering, Shenzhen University, Shenzhen 518060, China\\
$^2$Laboratory for Optics and Optoelectronics, Department of Physics
and Electronic Information Science, Hengyang Normal University,
Hengyang 421002, China\\}
\date{\today}

\begin{abstract}
We propose that the full Poincar\'{e} beam with any polarization
geometries can be pictorially described by the hybrid-order
Poincar\'{e} sphere whose eigenstates are defined as a
fundamental-mode Gaussian beam and a Laguerre-Gauss beam. A robust
and efficient Sagnac interferometer is established to generate any
desired full Poincar\'{e} beam on the hybrid-order Poincar\'{e}
sphere, via modulating the incident state of polarization. Our
research may provide an alternative way for describing the full
Poincar\'{e} beam and an effective method to manipulate the
polarization of light.

\end{abstract}


\maketitle

\section{Introduction}
Polarization is a fundamental property of light. Conventional states
of polarization, including linear, circular, and elliptical
polarizations have homogeneous spatial distribution. Arbitrary
polarization state can be described pictorially as a point in the
surface of a unit sphere, named as Poincar\'{e}
sphere~\cite{Born1997}. In recent years, optical beams with
inhomogeneous polarization in the transverse plane, such as
cylindrical vector beam, have drawn much attention due to its
intriguing properties under high-numerical-aperture focusing and a
range of potential applications, such as in optical trapping,
high-resolution metrology, and electron
acceleration~\cite{Hall1996,Zhan2009}. This kind of beams can be
represented by higher-order Poincar\'{e}
spheres~\cite{Holleczek2011,Milione2011}. In fact, optical beams can
simultaneously exhibit inhomogeneous polarization and phase, such as
cylindrical vector vortex
beams~\cite{Niv2006,Beresna2011,Chen2011,Zhao2013,Yi2014}. Apart
from the inhomogeneous polarization, the vector vortex beams have a
helical (vortex) phase factor. We have generalized the fundamental
Poincar\'{e} sphere to a hybrid-order Poincar\'{e} sphere to
describe the vector vortex beams\cite{Yi2015}. Recently, a new class
of beam with inhomogeneous polarization distribution, full
Poincar\'{e} beam (FPB), which covers all possible polarization
states over the fundamental Poincar\'{e} sphere in its transverse
plane, has been proposed~\cite{Beckley2010,Beckley2012}. It can be
generated by coaxial superposition of a fundamental-mode Gaussian
beam and a Laguerre-Gauss
beam~\cite{Beckley2010,Beckley2012,Han2011,Galvez2012} or untuned
q-plates~\cite{Cardano2013,Bauer2015}. The FPBs exhibit polarization
singularities and therefore have great potential in the singular
optics~\cite{Cardano2013,Bauer2015,Philip2012,Kumar2014,Galvez2014}.
In addition, the FPBs can be used for beam shaping~\cite{Han2011} or
studying the geometric Pancharatnam-Berry in optical
system~\cite{Vega2011,Kumar2013}.

In this paper, we propose that the FPB can be pictorially described
by the hybrid-order Poincar\'{e} sphere which is established by
defining the orthogonal eigenstates with a fundamental Gaussian beam
and a Laguerre-Gauss beam. A Sagnac interferometer is employed to
generate the FPB. The setup composed of a spatial light modulator, a
polarizing beam splitter, and a mirror. The setup converts the
horizontal polarization light into a vortex-bearing beam and remains
the vertical one unchanged. Their coaxial superposition manifests as
a FPBs. Controlling the incident polarization with a polarizer and a
quarter-wave plate, we can realize arbitrary FPBs on the
hybrid-order Poincar\'{e} sphere.

\section{Hybrid-order Poincar\'{e} sphere}
The fundamental Poincar\'{e} sphere can represent all the
conventional states of polarization, linear, circular, and
elliptical. The north and south poles correspond to the left- and
right-handed circular polarization (eigenstates), respectively.
Their equal-weight superposition results in the states of
polarization on the equator, i.e., linear polarizations. Others
located between the poles and the equator, are elliptical
polarizations. While the orthogonal eigenstates of the hybrid-order
Poincar\'{e} sphere are defined as two circular polarizations with
different vortex phase factors, and they can synthetise the vector
vortex beams~\cite{Yi2015}.

\begin{figure}
\centerline{\includegraphics[width=10cm]{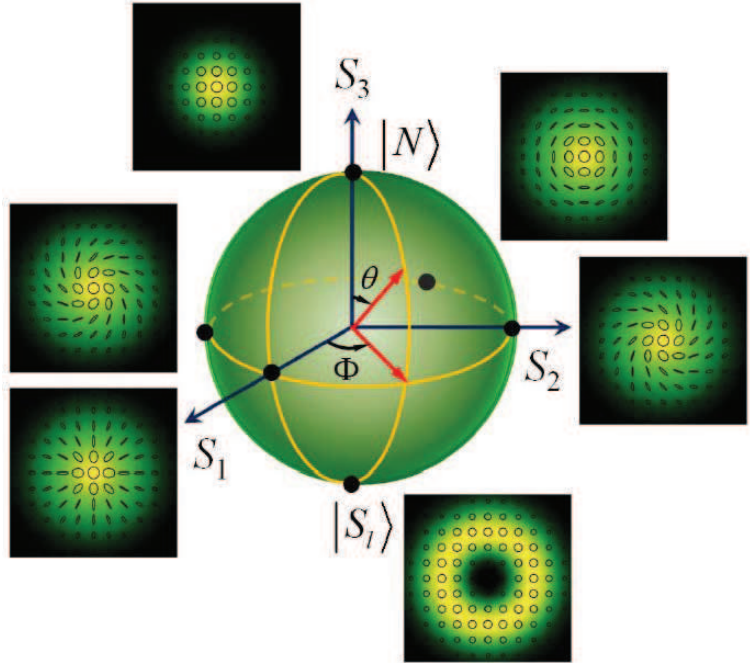}}
\caption{\label{Fig1} Schematic of the hybrid-order Poincar\'{e}
sphere for $l = 2$. The insets indicate the intensity background and
polarization geometries (ellipses) of six FPBs on the equator and
poles, respectively.}
\end{figure}

We here establish a hybrid-order Poincar\'{e} sphere with its
orthogonal circular-polarization eigenstates being a
fundamental-mode Gaussian beam and a Laguerre-Gauss beam. Suitable
superposition of the eigenstates can generate any desired FPBs, and
thus the proposed hybrid-order Poincar\'{e} sphere can be used to
describe the FPB, i.e., any FPB corresponds to a point on the
surface of the unit sphere. An alternative algebraic representation
of the FPB can be written as
\begin{equation}
\left|\psi_{l}\right\rangle={\psi _N}\left| N
\right\rangle+\psi_S^l\left|{{S_l}}\right\rangle,
\end{equation}
where the circular polarization eigenstates are
\begin{equation}
\left| N \right\rangle  = \frac{1}{{\sqrt 2 }}({\hat e_x} + i{\hat
e_y}){\rm{L}}{{\rm{G}}_{0,0}},~\left| {{S_l}} \right\rangle  =
\frac{1}{{\sqrt 2 }}({\hat e_x} - i{\hat
e_y}){\rm{L}}{{\rm{G}}_{0,l}}.
\end{equation}
Here, \^{e}$_x$ (\^{e}$_y$) is the unit vector in the $x$ ($y$)
direction. $\psi_N$ and $\psi_S^l$ are complex amplitudes of the
orthogonal circular polarization bases, respectively.
LG$_{0,0}(r,z)$ represents a fundamental-mode Gaussian beam
propagating in the $z$ direction with $r =(x^2+y^2)^{1/2}$ being the
transverse coordinate, which can be expressed as
\begin{equation}
{\rm{L}}{{\rm{G}}_{0,0}}(r,z) = {A_0}\frac{{{w_0}}}{{w(z)}}\exp
\left[ { - \frac{{{r^2}}}{{{w^2}(z)}}} \right]\exp \left[ { -
i\left( {kz - \frac{{k{r^2}}}{{2R(z)}} - \xi (z)} \right)} \right],
\end{equation}
where $A_0$ is a constant, $k$ is the free-space wavevector,
$w_0=w(0)$ is the beam waist size, $R(z)$ is the radius of curvature
of the beam's wavefronts, and $\xi(z)=\tan^{-1}[z/z_R]$ is the Gouy
phase with $z_R=kw_0^2/2$ being the Rayleigh distance. The
Laguerre-Gauss beam LG$_{0,l}$ can be expressed as
\begin{equation}
{\rm{L}}{{\rm{G}}_{0,l}}(r,z) = {\rm{L}}{{\rm{G}}_{0,0}}(r,z){\left[
{\frac{{\sqrt 2 r}}{{w(z)}}} \right]^{|l|}}\exp (il\varphi )\exp
\left[ {i|l|\xi (z)} \right],
\end{equation}
where $\varphi=\arctan(y/x)$ is the azimuthal angle and $l$ is the
topological charge of the vortex phase.

The state of polarization on the hybrid-order Poincar\'{e} sphere
can be defined by two independent parameters $(\theta,\Phi)$ with
$\theta\in[0,\pi]$ and $\Phi\in[0,2\pi]$, which fix the colatitude
and azimuthal angles on the sphere, as depicted in Fig. 1. Then we
can obtain that
\begin{equation}
\tan\left({\frac{\theta}{2}}\right)= \frac{{\left| {\psi _S^l}
\right|}}{{\left| {\psi _N} \right|}},~\Phi=\arg(\psi_S^l)-\arg
(\psi _N).
\end{equation}

For $\theta = 0$ and $\pi$, Equation (1) represents the FPBs on the
north and south poles, respectively. When $\theta=\pi/2$, i.e.,
$|\psi_N|=|\psi_S^l|$, the FPB locates on the equator, representing
the equal-intensity superposition of the two orthogonal bases. Other
values of $\theta$ indicate the FPBs on the spheres between the
poles and the equator. In addition, $\Phi$ defines the value of the
longitude.

The sphere's Cartesian coordinates now can be represented by the
Stokes parameters~\cite{Born1997}
\begin{equation}
S_0=|\psi _N|^2+|\psi_S^l|^2,~S_1=2|\psi
_N||\psi_S^l|\cos\Phi,~S_2=2|\psi _N||\psi_S^l|\sin\Phi,~S_3=|\psi
_N|^2-|\psi_S^l|^2.
\end{equation}
$S_0$ is unit radius of the hybrid-order Poincar\'{e} sphere and
$S_{1,2,3}$ are the sphere's Cartesian coordinates. For $l = 0$, the
hybrid-order Poincar\'{e} sphere reduces to the well-known
fundamental Poincar\'{e} sphere. Figure 1 depicts a hybrid-order
Poincar\'{e} sphere for $l=2$, where six FPBs located on the equator
and poles are given.

\section{Experimental scheme}
We implement an experimental setup to generate the FPBs, as depicted
in Fig. 2. A Glan laser polarizer (GLP) and quarter-wave plate
(QWP1) can convert the laser beam output from the He-Ne laser
(operational wavelength $\lambda = 632.8$ nm and beam waist size
$w_0=0.7$ mm) into any desired state of polarization (linear,
circular, and elliptical). Assumed the transmission axis of the GLP
and optical axis of the QWP1 incline angles $\gamma$ and $\psi$ with
respect to the horizontal direction, respectively, the Jones vector
of the light beam after the QWP1 can be written as
\begin{equation}
[\cos \gamma - i\cos (2\psi - \gamma )]{\rm{L}}{{\rm{G}}_{0,0}}{\hat
e_x} + [\sin \gamma - i\sin (2\psi - \gamma
)]{\rm{L}}{{\rm{G}}_{0,0}}(r,z){\hat e_y}.
\end{equation}

Then the beam propagates into a Sagnac interferometer consisting of
a polarizing beam splitter (PBS), a phase-only spatial light
modulator (SLM), and a mirror (M). The PBS transmits the horizontal
polarization component and reflects the vertical polarization
component of the incident beam. The horizontal polarization sub-beam
acquires a helical vortex phase after reflecting by the SLM and
propagates anticlockwise in the interferometer, which is a good
approximation of the vortex-bearing Laguerre-Gauss beam LG$_{0,l}$,
expressed by Eq. (4). Note that the reflection of the mirror
reverses the topological charge of the vortex from $l$ to $-l$.
While the vertically polarized sub-beam propagates clockwise and do
not acquire any spatially inhomogeneous phase. Because the SLM only
offers phase modulation to the horizontal polarization beam and
reflects off the vertical one without vortex phase modulation. The
Jones vector of the beam before the QWP2 is
\begin{equation}
[\cos\gamma-i\cos(2\psi-\gamma)]{\rm{L}}{{\rm{G}}_{0,l}}{\hat
e_x}+[\sin\gamma-i\sin(2\psi-\gamma )]{\rm{L}}{{\rm{G}}_{0,0}}{\hat
e_y}.
\end{equation}

\begin{figure}
\centerline{\includegraphics[width=10cm]{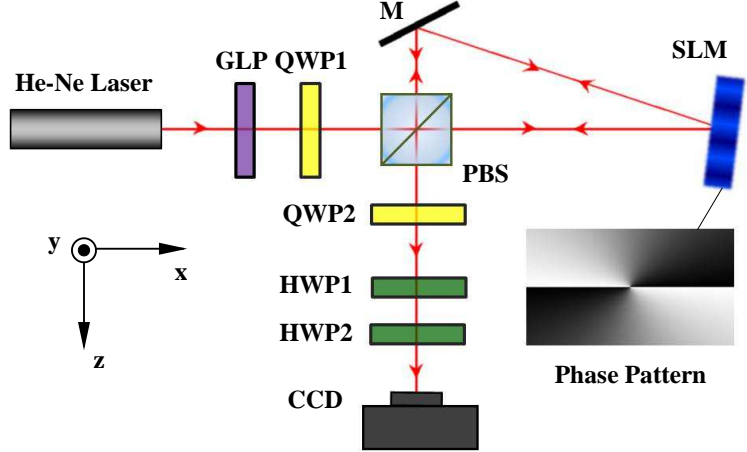}}
\caption{\label{Fig2} Experimental setup for generating the FPBs. An
He-Ne laser outputs a linearly polarized light beam whose
polarization state can be tuned to arbitrary linear, circular, or
elliptical polarization by modulating the Glan laser polarizer (GLP)
and the quarter-wave plate (QWP1). Then the beam passes through the
PBS and is split into two sub-beams with the transmission beam being
a horizontal polarization and the reflection beam being a vertical
one. The spatial light modulator (SLM) applies a helical phase to
the horizontal component and reflects off the vertical component
without vortex phase modulation. The PBS, the SLM, and the Mirror
(M) forms a Sagnac interferometer. Another quarter-wave plate (QWP2)
converts the two sub-beams into orthogonal circular polarizations
located on the north and south poles on the hybrid-order
Poincar\'{e} sphere. Two cascaded half-wave plate are employed to
manipulate the FPBs moving along the latitude of the sphere. A CCD
camera is used to record the beam intensity. The insets: a phase
pattern with topological charge $l=2$ displayed on the SLM.}
\end{figure}

Then the two components pass through another quarter-wave plate
(QWP2) with its optical axis inclining $45^\circ$ to the horizontal
direction which converts them into two orthogonal circular
polarizations located on the north and south of the hybrid-order
Poincar\'{e} sphere. As the Jones matrix of the quarter-wave plate
can be expressed as
\begin{equation}
\left( \begin{array}{l}
 ~1 \\
  - i \\
 \end{array} \right.\left. \begin{array}{l}
  - i \\
  ~1 \\
 \end{array} \right).
\end{equation}
Hence, the Jones vector of the optical beam after the QWP2, i.e.,
the FPB based on the circular polarization eigenstates, is given by
\begin{equation}
[- i\sin \gamma - \sin (2\psi -
\gamma)]{\rm{L}}{{\rm{G}}_{0,0}}({\hat e_x} + i{\hat e_y}) + [\cos
\gamma - i\cos (2\psi  - \gamma )]{\rm{L}}{{\rm{G}}_{0,l}}({\hat
e_x} - i{\hat e_y}).
\end{equation}
It represents the superposition of two orthogonal circular
polarizations located on the north and south poles of the
hybrid-order Poincar\'{e} sphere, respectively. So, we can obtain
that
\begin{equation}
{\psi _N} = [ - i\sin \gamma  - \sin (2\psi - \gamma )],~\psi _S^l =
[\cos \gamma - i\cos (2\psi-\gamma)].
\end{equation}
Substituting the above two equations into Eq. (5), we can establish
the relationship between the generated FPBs and the two tunable
parameters, $\gamma$ and $\psi$, i.e.,
\begin{equation}
\tan\left(\frac{\theta}{2}\right)=\left[\frac{{{{\cos }^2}\gamma +
{{\cos }^2}(2\psi - \gamma )}}{{{{\sin }^2}\gamma  + {{\sin
}^2}(2\psi - \gamma )}}\right]^{1/2},~\Phi ={\tan ^{ - 1}}\left[ -
{\frac{{\cos (2\psi - \gamma )}}{{\cos \gamma}}}\right]-{\tan ^{ -
1}}\left[ {\frac{{\sin \gamma }}{{\sin(2\psi -\gamma )}}}\right].
\end{equation}
In this case, any desired FPB can be generated over the hybrid-order
Poincar\'{e} sphere by suitably modulating the GLP and QWP1.

In fact, LG$_{0,l}$ beam exhibits an additional phase factor
$\exp[i|l|\xi(z)]$ which is $z$-dependent, that is, the polarization
geometry of the generated FPB changes upon propagation. To
compensate this phase, we employ two cascaded half-wave plates (HWP1
and HWP2) with their optical axes inclined angles $\zeta_1$ and
$\zeta_2$ with respect to the horizontal direction, respectively.
Then the Jones vector of the output beam is given by
\begin{equation}
\psi_N\exp[ - i2({\zeta _1} - {\zeta
_2})]{\rm{L}}{{\rm{G}}_{0,0}}(r,z)({\hat e_x} + i{\hat e_y}) +
\psi_S^l\exp [i2({\zeta _1} - {\zeta
_2})]{\rm{L}}{{\rm{G}}_{0,l}}(r,z)({\hat e_x} - i{\hat e_y}).
\end{equation}

One finds that the two cascaded half-wave plates only apply an
opposite phase factor $\exp[\pm i2(\zeta_1-\zeta_2)]$ determined by
the relative direction of optical axes of the two plates, to the two
circular polarization eigenstates. From Eq. (3), one knows this only
changes $\Phi$. And $\theta$ remains unchanged. So, the two cascaded
half-wave plates will provide an effective way to manipulate the FPB
moving along the latitudes of the hybrid-order Poincar\'{e} sphere.

\section{Results and discussion}
We have two degrees of freedom to manipulate the generated FPBs. For
one thing, we can modulate the topological charge $l$ of the
sub-beams produced by the SLM via switching the phase picture
displayed on the SLM. For different topological charges, there
exists different hybrid-order Poincar\'{e} spheres. For another, we
can obtain any desired FPBs on the hybrid-order Poincar\'{e} sphere
by modulating the transmission axis direction of the GLP and the
optical axis direction of the QWP1, i.e., tuning the incident state
of polarization.

\begin{figure}
\centerline{\includegraphics[width=14cm]{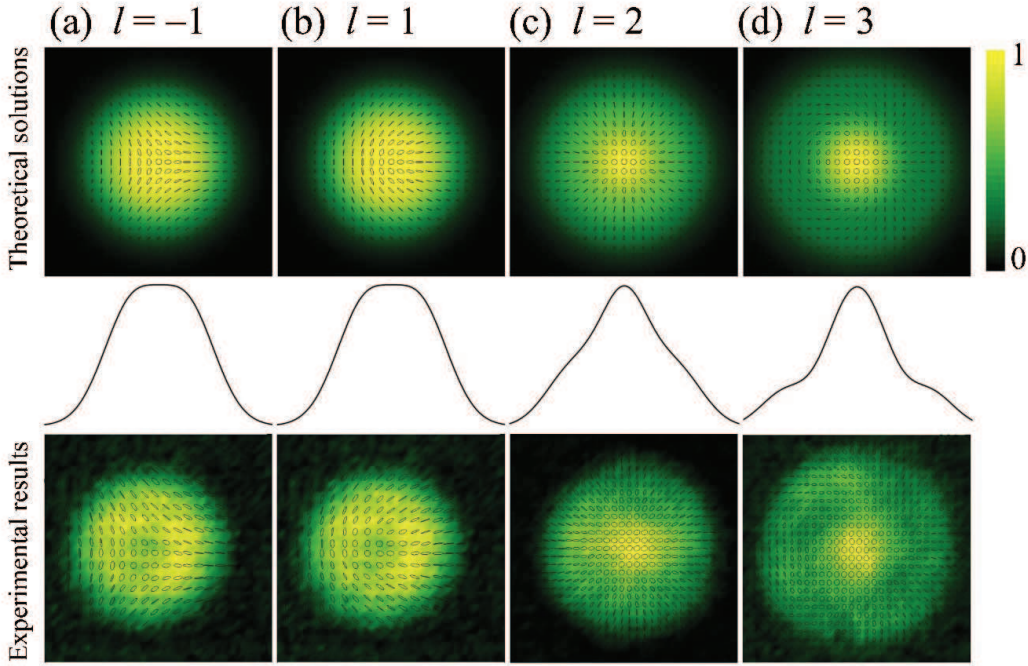}}
\caption{\label{Fig3} Maps of the polarization states (ellipses) on
the background of the intensity distribution of the FPBs for
different topological charges ($l=-1$, 1, 2, and 3) under the
condition of $(\theta,\Phi)=(90^\circ,0)$. The top row depicts the
theoretical solutions of the FPBs and their polarization geometries.
The middle row gives the one-dimensional intensity distribution of
the FPBs at $y=0$. The bottom row is the corresponding experimental
results.}
\end{figure}

We first generate the FPBs for the cases of $l=-1$, 1, 2, and 3, as
depicted in Fig. 3. Different topological charges $l$ corresponds to
different hybrid-order Poincar\'{e} sphere. We set $\gamma = \pi/4$
and $\psi=0$, so the FPBs locate on the sphere with
$(\theta,\Phi)=(\pi/2, 0)$, i.e., at the $S_1$ on the equator. To
retrieve the local polarization distribution, we measure the Stokes
parameters by a typical setup, a quarter-wave plate, a polarizer
(not shown in the figure), and a CCD. So, the polarization state can
be reconstructed pixel by pixel via solving the measured Stokes
parameters. The experimental results agree with the theoretical
solutions. Note that these FPBs have flattop intensity distribution,
especially for $l=\pm1$, although the polarization state is
inhomogeneous. The Gaussian distribution of intensity for one
eigenstate exactly offsets the doughnut-shaped LG$_{0,l}$ intensity
of another eigenstate.

The local polarization state of the FPB is determined by the local
amplitude and phase of the two eigenstates. Because of the circular
symmetry of beam intensity in the azimuthal direction, the local
amplitude ratio of the two eigenstates remains unchanged and the
phase changes continuously. So the local polarization ellipse keep
the same ellipticity, but varies the direction of principal axes.
Obviously, the polarization pattern has a topological charge of
$l/2$. In the radial direction, the local amplitude ratio varies
continuously but the phase difference is a constant, so the
ellipticity of the local polarization changes from left circular,
through elliptical, to linear, and then to right-handed elliptical,
and the direction of principal axes remains unvaried. When the
amplitude ratio equal to 1, the local polarization is linear, which
is associated with the L-lines
singularites~\cite{Cardano2013,Kumar2014}. In a word, the
polarization geometry of each FPB covers all the possible state on
the fundamental Poincar\'{e} sphere.

\begin{figure}
\centerline{\includegraphics[width=11cm]{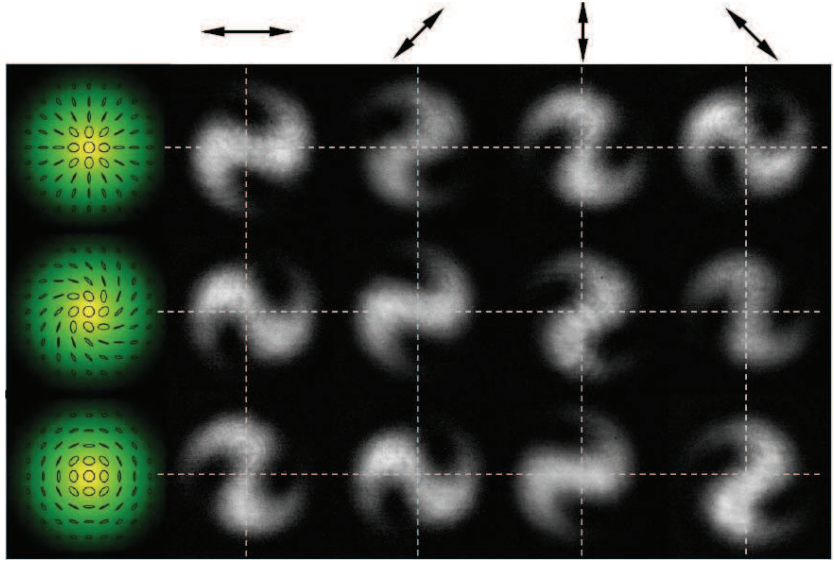}}
\caption{\label{Fig4} CCD recorded intensity of the FPBs on the
equator ($\Phi=0$, $\pi/2$, and $\pi$) after a polarizer (not shown
in Fig.3) with different direction of transmission axis (represented
by the arrows in the first row). }
\end{figure}

In fact, the LG$_{0,l}$ beam exhibits a $z$-dependent phase factor,
$\exp[i|l|\xi(z)]$, which results in the variation of $\Phi$ upon
beam propagation ($\theta$ remains unchanged). In other words, the
FPBs will move along a latitude on the hybrid-order Poincar\'{e}
sphere when they propagate in free space. To compensate this
additional phase introduced by beam propagation, two cascaded
half-wave plates, HWP1 and HWP2, are employed. Because the half-wave
plates only influence the phase of the eigenstates, as depicted
theoretically in Eq. (13). We choose three typical FPBs on the
equator of the hybrid-order Poincar\'{e} sphere ($\Phi=0$, $\pi/2$,
and $\pi$) and analyze their polarization using a polarizer, as
depicted in Fig. 4. Rotating the polarizer, the recorded intensities
in the CCD camera shows ``S''-shaped patterns whose rotation
direction is the same as that of the polarizer. At a propagation
distance ($z\neq0$), due to the beam propagation, $\Phi$ changes
continuously. By modulating the relative optical axis direction of
the HWP1 and HWP2, i.e., $\zeta_1-\zeta_2$, and rotating the
polarizer, the intensity rotations depicted by Fig. 4 can be found
in this process. Take any case of the three states as a reference,
we can realized any other intermediate state on the equator.
Further, tuning the GLP and QWP1, the FPBs can be switched to any
desired latitude. Hence, any desired FPBs on the hybrid-order
Poincar\'{e} sphere can be realized by the combined use of a
polarizer and a quarter-wave plate controlling the incident state of
polarization and two cascaded half-wave plates manipulating the
phase of the eigenstates.

\begin{figure}
\centerline{\includegraphics[width=14cm]{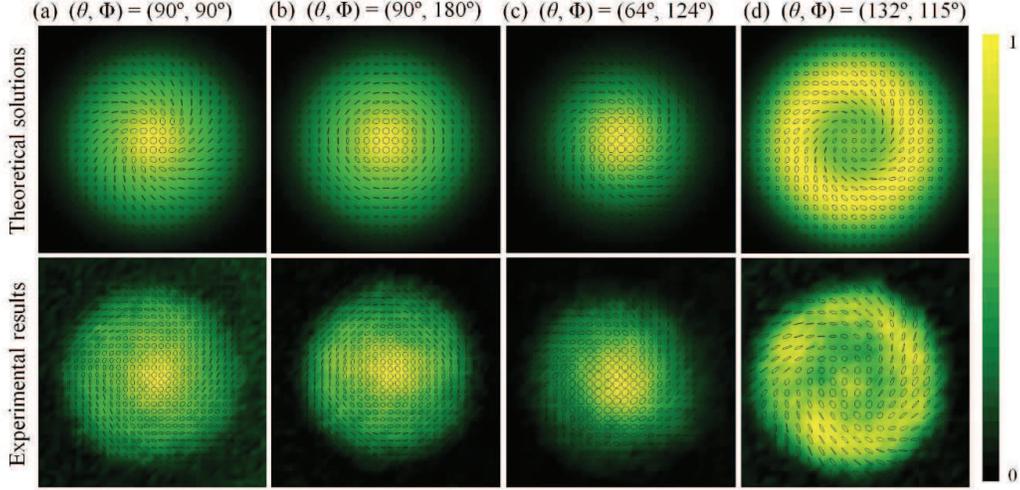}}
\caption{\label{Fig5} Examples of the FBPs on the hybrid-order
Poincar\'{e} sphere. (a) $(\theta,\Phi)=(90^\circ,90^\circ)$. (b)
$(\theta,\Phi) = (90^\circ,180^\circ)$. (c) $(\theta,\Phi) =
(64^\circ,124^\circ)$. (d) $(\theta,\Phi) = (132^\circ,115^\circ)$.
The top row is the theoretical solutions and the bottom row is the
corresponding experimental results. }
\end{figure}

By solving Eq. (13), we obtain the corresponding mapping
relationship between $(\theta,\Phi)$ and $(\gamma,\psi)$, so the
latitude of the FPB is first determined. Based on any case in Fig. 4
and tuning $\zeta_1-\zeta_2$ to the appropriate value, we can
therefore acquire the required FPBs on the hybrid-order Poincar\'{e}
sphere. FPBs located on $(\theta,\Phi)=(90^\circ,90^\circ)$,
$(90^\circ,180^\circ)$, $(64^\circ,124^\circ)$, and
$(132^\circ,115^\circ)$, are solved, as depicted in Fig. 5. The
first and the second FPBs are on the equator, and the third FPB
locates on the northern hemisphere, while the forth one is on the
southern hemisphere.

\section{Conclusions}
We have employed a hybrid-order Poincar\'{e} sphere to describe the
FPBs, whose eigenstates are a pair of orthogonally circularly
polarized Gaussian beam and Laguerre-Gaussian beam. An experimental
setup has also been established to generate any desired FPBs over
the hybrid-order Poincar\'{e} sphere. The combination usage of the
GLP, QWP1, HWP1, and HWP2 is crucial for manipulating the FPBs. It
is worth to mention that if the WQP2 is removed, the
linear-polarization based FPBs can be generated~\cite{Beckley2010}
and the corresponding hybrid-order Poincar\'{e} sphere with its
orthogonal eigenstates being linearly polarized Gaussian beam and
Laguerre-Gaussian beam can also be established. Our research may
provide an alternative way for describing the full Poincar\'{e} beam
and an effective method to study the polarization singularities and
geometric Panacharatnam-Berry phase of
light~\cite{Yi2015,Cardano2013,Vega2011,Kumar2013}.

\begin{acknowledgements}
This research was partially supported by the National Natural
Science Foundation of China (Grants No. 11447010), the Natural
Science Foundation of Hunan Province (Grant No. 2015JJ3026), the
China Postdoctoral Science Foundation (Grant No. 2014M562198), the
Science Foundation of Hengyang Normal University (Grant No. 12B37),
and the Construct Program of the Key Discipline in Hunan Province.
\end{acknowledgements}

\end{document}